\documentclass[prd,amsmath,amssymb,floatfix,nofootinbib,superscriptaddress,twocolumn,letterpaper]{revtex4}

\usepackage{aas_macros}

\usepackage[breaklinks, colorlinks, citecolor=blue, colorlinks=true, linkcolor=blue, urlcolor=blue]{hyperref}
\usepackage{ifthen}

\def\eprinttmp@#1arXiv:#2 [#3]#4@{
\ifthenelse{\equal{#3}{x}}{\href{http://arxiv.org/abs/#1}{#1}}{\href{http://arxiv.org/abs/#2}{arXiv:#2} [#3]}}
\providecommand{\eprint}[1]{\eprinttmp@#1arXiv: [x]@}
\newcommand{\adsurl}[1]{\href{#1}{ADS}}

\usepackage{amsmath}
\usepackage{natbib}
\usepackage{multirow}
\usepackage{subfigure}
\usepackage{float}
\usepackage{epstopdf}
\usepackage{color}
\usepackage[usenames,dvipsnames]{xcolor}
\usepackage{tikz}

\def\L2{\ifmmode L_2\else $L_2$\fi}
\def\Msolar{\ifmmode{\rm M}_{\mathord\odot}\else${\rm M}_{\mathord\odot}$\fi}
\newcommand{\prof}{{\cal Y}}
\newcommand{\muKarcmin}{\mu {\rm K}\, {\rm arcmin}}
\newcommand{\hn}{\hat{n}}
\newcommand{\bn}{\hat{\textbf{n}}}

\newcommand{\bl}{\vec{\ell}}
\newcommand{\bL}{\vec{L}}
\newcommand{\be}{\begin{equation}}
\newcommand{\ee}{\end{equation}}

\newcommand{\tcl}{\hat{\cal C}}
\newcommand{\mre}{\phi^{BH}} %\Gamma}
\newcommand{\mix}{M}
\newcommand{\tlen}{T}
\newcommand{\tobs}{T^{\rm obs}}
\newcommand{\comments}[1]{}

\newcommand{\threej}[6]{\left(
    \begin{array}{ccc}
        \! #1\! & #2\!  & #3\!  \\
        \! #4\! & #5\!  & #6\!
      \end{array}
    \right)}
\newcommand{\sixj}[6]{\left\{
                      \begin{array}{ccc}
    #1 & #2  & #3  \\
    #4 & #5  & #6
                      \end{array}
               \right\}}
\newcommand{\Planck}{Planck}

\begin{document}

\title{Extragalactic Foreground Contamination \\ in Temperature-based CMB Lens Reconstruction}

\author{Stephen J. Osborne}
\affiliation{Stanford University, 382 Via Pueblo, Varian Building, Stanford, CA, 94305-4060}
\affiliation{Kavli Institute for Particle Astrophysics and Cosmology at Stanford University, 452 Lomita Mall, Stanford, CA, 94305-4085}

\author{Duncan Hanson}
\affiliation{Department of Physics, McGill University, 3600 rue University, Montr\'eal, QC, Canada H3A 2T8}

\author{Olivier Dor\'e}
\affiliation{Jet Propulsion Laboratory, California Institute of Technology, 4800 Oak Grove Drive, Pasadena CA 91109, U. S. A.}
\affiliation{California Institute of Technology, Pasadena, California, U.~S.~A.}

\begin{abstract}
We discuss the effect of unresolved point source contamination on estimates of the CMB lensing potential, 
from components such as the thermal Sunyaev-Zel'dovich effect, radio point sources, and the Cosmic Infrared Background.
We classify the possible trispectra associated with such source populations, and
construct estimators for the amplitude and scale-dependence of several of the major trispectra. 
We show how to propagate analytical models for these source trispectra to biases for lensing.
We also construct a ``source-hardened'' lensing estimator which experiences significantly smaller biases when exposed to unresolved point sources than the standard quadratic lensing estimator.
We demonstrate these ideas in practice using the sky simulations of Sehgal et. al.,
for cosmic-variance limited experiments designed to mimic ACT, SPT, and {\it Planck}.
\end{abstract}

\maketitle

\section{Introduction}
\label{sec:intro}

Gravitational lensing of the cosmic microwave background (CMB) by large-scale 
structure is a long predicted effect \citep{1987A&A...184....1B}, which has only recently
become detectable \citep{2007PhRvD..76d3510S,2008PhRvD..78d3520H,2011PhRvL.107b1301D}. 
Modern high-resolution CMB measurements from ACT, SPT, and the \Planck\ satellite are now able to
measure the power spectrum of the CMB lensing potential at high significance (between $4\rightarrow25\sigma$),
and have successfully used it for cosmological parameter
fitting~\citep{2011PhRvL.107b1301D,vanEngelen:2012va,Ade:2013mta}.

The power spectrum of the CMB lensing potential is not directly measurable,
but must be inferred from the non-Gaussian statistical properties that it induces in the
observed CMB. The trispectrum, or connected 4-point function, of the CMB
is very sensitive to lensing \citep{Hu:2001fa} 
and is the basis for all current estimators of the lensing potential power spectrum 
\citep[e.g.][]{2003PhRvD..67h3002O,2003PhRvD..67d3001H,Bucher:2010iv}.
A potential source of error in these measurements can be caused by additional sources of non-Gaussian signal
in the observed sky, such as extragalactic foreground contamination from radio/infrared point sources 
and the Sunyaev-Zel'dovich effect, which may be misinterpreted as a lensing signal and provide a
source of bias for lensing estimates.

In this paper, we classify the main terms in the CMB and foreground point-source trispectra,
construct estimators which can be used to measure their amplitude, and discuss how to propagate them 
to biases for lensing power spectrum measurements. 
Furthermore, we develop modified lensing estimators which are less sensitive to point source contamination. 
Components of this methodology have been used already in the analysis of \cite{Ade:2013mta}. 
This paper represents a more detailed discussion and analysis of the estimator-based approach.  

The bias to lens reconstruction from extragalactic foregrounds has been studied before.
The most detailed analyses have focussed on the contamination of the cross-correlation of the CMB
lensing potential with external tracers.
Unresolved source contamination was found to provide
the largest source of systematic error for cross-correlation of the CMB lensing potential with NVSS
quasars \citep{2007PhRvD..76d3510S,2008PhRvD..78d3520H}. 
The cross-correlation between the SPT-measured CMB lensing potential and the Cosmic Infrared Background (CIB)
as probed by Herschel was estimated to have a bias of a few percent~\citep{Holder:2013hqu}.
Although useful as a guide to the magnitude of lensing bias effects, these calculations concern the bispectrum (3-point) correlation
between the CMB and external tracers, and are only a subset of the full set of trispectrum (4-point) terms of concern for 
lensing auto-spectrum measurements.
The trispectrum has received relatively less attention.
In the case of polarization lens reconstruction, the complete set of trispectrum terms can be calculated under the
assumption that polarization angles from different sources are uncorrelated~\citep{2008arXiv0811.3916S}.
Under this assumption the contamination from
unresolved radio point sources in polarization was found to be small, however this situation does not necessarily carry 
over to temperature, where radio source emission is larger relative to the CMB fluctuations, and
Sunyaev-Zel'dovich (SZ) and infrared source emission are also a larger concern. 
In the case of temperature, published analyses have used the simulations of~\citet{2010ApJ...709..920S} to investigate
numerically the magnitude of source contamination, finding percent-level biases 
\citep{2010arXiv1007.3519C,2011PhRvL.107b1301D,vanEngelen:2012va}.
In this work, we will also use the simulations of~\cite{2010ApJ...709..920S} to test several aspects of our results. 

During the preparation of this work, we became aware of a new paper by van Engelen et. al. \cite{vanEngelen2013}, which 
performs a thorough analysis of CMB lensing biases from extragalactic foregrounds,  
incorporating improvements in our understanding of the sky given by the wealth of new sub-mm data from Herschel, ACT, and SPT and \Planck. 
The results of van Engelen et. al. and this work are very complementary, 
with the former focussing on the range of biases in the allowed space of models, 
and the latter focussing on estimators for the foreground trispectra and methods to mitigate them.

Throughout this work, the following simple model of point sources will provide useful intuition.
Consider laying down a field of density contrast $1 + \delta (\vec{x})$ throughout the Universe.
Then proceed to populate this Universe with point sources such that the source number density traces
the density perturbations. 
On sufficiently large scales, the distribution function from which these sources are drawn 
can be approximated as a linear function of the density, so that e.g. a region of space where
$\delta(\vec{x})$ is twice as large will have (on average) twice as many sources.
If we consider the density field $\delta(\vec{x})$ to be fixed, and look at multiple realizations of
the source distribution, then the 
``shot noise'' from realization-to-realization has statistical properties that are similar to
instrumental noise, but with the
noise level modulated by the large-scale density field rather than by integration time.
This modulation introduces a source of statistical anisotropy, with properties that
can be studied in an analogous way to the statistical anisotropy that lensing introduces.
Of course, in reality we do not have multiple realizations of the Universe with the same density
contrast. Both point sources and lensing ultimately appear in CMB maps as a source of trispectrum 
non-Gaussianity, rather than statistical anisotropy. 
Nevertheless, we find that this picture of point sources as a source of inhomogeneous noise
provides useful intuition.

The remainder of this paper is as follows.
In Sect.~\ref{sec:lens_recon} we review the procedure of lens reconstruction.
In Sect.~\ref{sec:source_trispectrum} we derive templates for the shape of extragalactic
	foreground contributions to the trispectrum, and discuss how to propagate several of them to biases
	for lensing estimators.
In Sect.~\ref{sec:triest} we discuss how to construct estimators for the point source trispectra, as well as lensing
	estimators which are less sensitive to them.
In Sect.~\ref{sec:fore_sim} we demonstrate this approach using the simulations of~\citet{2010ApJ...709..920S}.
We conclude in Sect.~\ref{sec:conc}.

\section{Lens Reconstruction}
\label{sec:lens_recon}

\subsection{Lensing Potential}

In this section, we briefly review the process we use to estimate the lensing potential power spectrum $C_L^{\phi\phi}$
from temperature maps of the microwave background, using a formalism which will make the discussion of the
proceeding sections more compact.
Lensing is a remapping operation, described in real space by
\be
T(\hat{n}) = \tilde{T}(\hat{n} + \nabla \phi(\hat{n})),
\ee
where $\tilde{T}$ is the primordial, unlensed CMB temperature, and $\phi$ is the CMB lensing potential.
We begin by considering the lensing potential $\phi$ to be fixed, 
and calculate the statistics of the CMB after averaging over realizations of the Gaussian $\tilde{T}(\hat{n})$.
Lensing is a linear operation, and so if the lensing potential is fixed the observed sky remains Gaussian, 
however it becomes statistically anisotropic and its covariance acquires off-diagonal elements. 
At first order in the lensing potential we have
\be
\langle T_{\ell_1 m_1} T_{\ell_2 m_2} \rangle = \sum_{LM} \threej{\ell_1}{\ell_2}{L}{m_1}{m_2}{M} W_{\ell_1 \ell_2 L}^{\phi} \phi_{LM},
\ee
where the ensemble average is taken over realizations of $\tilde{T}_{lm}$ and the lensing ``weight function'' is given by
\begin{multline}
W_{\ell_1 \ell_2 L}^{\phi} 
= -  \sqrt{\frac{(2\ell_1+1)(2\ell_2+1)(2L+1)}{4\pi}}  \, 
 \\ \times  C_{\ell_1}^{TT} \left( \frac{1 + (-1)^{\ell_1 + \ell_2 + L}}{2} \right) \threej{\ell_1}{\ell_2}{L}{1}{0}{-1} 
 \\ \times \sqrt{ L(L+1) \ell_1 (\ell_1+1) }+ (\ell_1 \leftrightarrow \ell_2).
\end{multline}
The statistical anisotropy introduced by lensing can be probed with quadratic estimators $\bar{x}$,
derived by maximizing the likelihood of the observed CMB \cite{2003PhRvD..67d3001H}, 
and constructed as
\be
\bar{x}_{LM} = \frac{1}{2} 
\sum_{\ell_1 m_1, \ell_2 m_2} 
\threej{\ell_1}{\ell_2}{L}{m_1}{m_2}{-M}
W^{x}_{\ell_1 \ell_2 L}
\bar{T}_{\ell_1 m_1} \bar{T}_{\ell_2 m_2},
\label{eqn:qe}
\ee
where $\bar{T}_{lm}$ are inverse-variance filtered CMB multipoles and $W^{x}_{\ell_1 \ell_2 L}$ is a
weight function for the quadratic estimator.
The weight function is usually taken to be a matched filter for lensing with $W^{x} = W^{\phi}$
and the estimator denoted as $\bar{\phi}$, however it can also be advantageous to use other weight functions
with reduced sensitivity to certain systematic effects \citep{Namikawa:2012pe}. 
We will show that this is also the case with point source contamination.
For a full-sky experiment with homogeneous instrumental noise, the optimal inverse-variance 
filter is given by $F_{\ell} = [C_\ell^{TT} + N_\ell^{TT}]^{-1}$, where $C_\ell^{TT}$ is a fiducial
CMB power spectrum and $N_\ell^{TT}$ is the (beam-deconvolved) instrumental noise and extragalactic foreground power spectrum.
This is applied to an observed (beam-convolved) data map $\tobs(\hat{n})$ as
\be
\bar{T}_{lm} = F_{\ell} B_\ell^{-1} \int d\hat{n} \,\, Y_{lm}^* \, \tobs(\hat{n}),
\label{eqn:tbar}
\ee
where $B_{\ell}$ is the instrumental beam transfer function.
The estimator $\bar{x}$ responds to $\phi$
such that averaging over CMB realizations with a fixed realization of $\phi$ gives
%\be
$
\left< \bar{x}_{LM} \right> = {\cal R}^{x \phi}_L \phi_{LM},
$
%\ee
where the response function ${\cal R}$ is given by
\be
{\cal R}^{x \phi}_L = \frac{1}{2L+1} \sum_{\ell_1 \ell_2} \frac{1}{2} W^{x}_{\ell_1 \ell_2 L} W^{\phi}_{\ell_1 \ell_2 L} F_{\ell_1} F_{\ell_2}.
\label{eqn:rxphi}
\ee
Estimates of the lensing potential may therefore be formed as
%\be
$\hat{\phi}^x_{LM} = ( {\cal R}^{x \phi}_L )^{-1} \bar{x}_{LM}$.
%\ee
The~\citet{2003PhRvD..67h3002O} lensing estimator, for example, is
%\be
$\hat{\phi}_{LM} = ( {\cal R}^{\phi \phi}_L )^{-1} \bar{\phi}_{LM}$,
%\ee
with $( {\cal R}^{\phi \phi}_L )^{-1}$ acting as the estimator normalization.
We will discuss this formalism further in Sect.~\ref{sec:triest}, giving
the response of additional estimators.

\subsection{Lensing Power Spectrum}
\label{sec:lensing_power_spectrum}

Estimates of the lensing potential power spectrum may be constructed straightforwardly from the potential estimates above.
The ensemble average (over realizations of both the primordial CMB and the lensing potential) is given explicitly by
\begin{multline}
\left< | \bar{x}_{LM}|^2 \right> = 
\frac{1}{4} (-1)^M
W^{x}_{\ell_1 \ell_2 L} W^{x}_{\ell_3 \ell_3 L} 
\sum_{\ell_1 m_1} \sum_{\ell_2 m_2} 
\sum_{\ell_3 m_3} \sum_{\ell_4 m_4}
\\ \times
\threej{\ell_1}{\ell_2}{L}{m_1}{m_2}{-M}
\threej{\ell_3}{\ell_4}{L}{m_3}{m_4}{M}
\\ \times
\langle \bar{T}_{\ell_1 m_1} \bar{T}_{\ell_2 m_2} \bar{T}_{\ell_3 m_3} \bar{T}_{\ell_4 m_4} \rangle.
\label{eqn:plm2}
\end{multline}
The ensemble-average of the 4-point function can be broken into connected (C) and disconnected (D) parts as
\begin{multline}
\langle \bar{T}_{\ell_1 m_1} \bar{T}_{\ell_2 m_2} \bar{T}_{\ell_3 m_3} \bar{T}_{\ell_4 m_4} \rangle = \\
\langle \bar{T}_{\ell_1 m_1} \bar{T}_{\ell_2 m_2} \bar{T}_{\ell_3 m_3} \bar{T}_{\ell_4 m_4} \rangle_C \\ +
\langle \bar{T}_{\ell_1 m_1} \bar{T}_{\ell_2 m_2} \bar{T}_{\ell_3 m_3} \bar{T}_{\ell_4 m_4} \rangle_D.
\end{multline}
The disconnected part is that which can be formed from the three Wick contractions of the four multipoles, and is given by
\begin{multline}
\langle \bar{T}_{\ell_1 m_1} \bar{T}_{\ell_2 m_2} \bar{T}_{\ell_3 m_3} \bar{T}_{\ell_4 m_4} \rangle_D = \\
\hspace{-3cm}
  \bar{C}_{\ell_1 m_1, \ell_2 m_2} \bar{C}_{\ell_3 m_3, \ell_4 m_4} \\ 
+ \bar{C}_{\ell_1 m_1, \ell_3 m_3} \bar{C}_{\ell_2 m_2, \ell_4 m_4} \\
+ \bar{C}_{\ell_1 m_1, \ell_4 m_4} \bar{C}_{\ell_2 m_2, \ell_3 m_3},
\end{multline}
where 
$C_{\ell_1 m_1, \ell_2 m_2} = \langle \bar{T}_{\ell_1 m_1} \bar{T}_{\ell_2 m_2} \rangle$ 
is the covariance matrix of $\bar{T}$.
We denote the contribution of the disconnected part to the ensemble average of Eq.~\eqref{eqn:plm2} as 
$N_{L}^{xx}$.

The connected part of the 4-point function is zero for purely Gaussian fluctuations, and so it
directly traces any non-Gaussianity in the map.
Following \cite{Hu:2001fa}, for a statistically isotropic non-Gaussian signal such as that due to lensing or extragalactic foregrounds 
the connected 4-point function
takes the form 
\begin{multline}
\langle {T}_{\ell_1 m_1} {T}_{\ell_2 m_2} {T}_{\ell_3 m_3} {T}_{\ell_4 m_4} \rangle_C 
= \sum_{LM} (-1)^{M} 
T^{\ell_1 \ell_2}_{\ell_3 \ell_4} (L) 
\\ \times
\threej{\ell_1}{\ell_2}{L}{m_1}{m_2}{-M}
\threej{\ell_3}{\ell_4}{L}{m_3}{m_4}{M},
\end{multline}
where $T^{\ell_1 \ell_2}_{\ell_3 \ell_4} (L)$ is known as the trispectrum.
Symmetry of the four multipoles imposes the requirement that the trispectrum may be written as
\begin{multline}
T^{\ell_1 \ell_2}_{\ell_3 \ell_4} (L)  = 
P^{\ell_1 \ell_2}_{\ell_3 \ell_4} (L) + 
(2L+1) \sum_{L'} \Bigg[ \\
(-1)^{\ell_2 + \ell_3} \sixj{\ell_1}{\ell_2}{L}{\ell_4}{\ell_3}{L'} P^{\ell_1 \ell_3}_{\ell_2 \ell_4} (L') \\
+ (-1)^{L + L'} \sixj{\ell_1}{\ell_2}{L}{\ell_4}{\ell_3}{L'} P^{\ell_1 \ell_4}_{\ell_3 \ell_2} (L') \Bigg].
\label{eqn:trispectrumencoding}
\end{multline}
The first term $P^{\ell_1 \ell_2}_{\ell_3 \ell_4} (L)$ is called the primary contraction of the trispectrum, 
while the second two terms are called the secondary contractions.
The primary contraction introduced by lensing, for example, is 
\be
{}^{\phi\phi}\! P^{\ell_1 \ell_2}_{\ell_3 \ell_4} (L) =
C_L^{\phi\phi} W^{\phi}_{\ell_1 \ell_2 L} W^{\phi}_{\ell_3 \ell_4 L}.
\ee
Given the discussion above, 
for an observed sky consisting of lensed CMB + Gaussian noise and
using the standard lensing estimator
%$\bar{x} \rightarrow \bar{\phi}$
with $W^x = W^{\phi}$, an estimate $\hat{C}_L^{\phi\phi}$ for the lensing potential power spectrum may be written implicitly as
\be
\hat{C}_L^{\phi\phi} + \mix^{L'}_{L} \hat{C}_{L'}^{\phi\phi} =  \left( \frac{1}{ {\cal R}^{\phi\phi} } \right)^2 \left[ \frac{1}{2L+1} \sum_{M} | \bar{\phi}_{LM}|^2 - N_{L}^{\phi \phi} \right],
\label{eq:clpp_est}
\ee
where the $L'$ index is summed over.
The mixing matrix $\mix^{L'}_{L}$ would be zero if there were only primary contractions of the trispectrum,
however it has small off-diagonal contributions due to the secondary contractions~\citep{Hu:2001fa}.
The mixing matrix may be inverted and used to obtain an estimate of $C_L^{\phi\phi}$
using Eq.~\eqref{eq:clpp_est}.
Alternatively, if a good approximation of the true power spectrum $C_L^{\phi\phi}$ is available,
the off-diagonal contributions can simply be subtracted from the LHS of Eq.~\eqref{eq:clpp_est}.
In this approach, this term is known as the ``$N^{(1)}$''  bias \citep{2003PhRvD..67l3507K}.

\section{Source Trispectra}
\label{sec:source_trispectrum}

In this section we outline the trispectrum configurations that can be generated by point sources. 
We model the source population as a collection of discrete sources $i$, with fluxes $S_i$, such
that the sky temperature in direction $\bn$ is given by
\be
\tilde{T}(\bn) = \tlen(\bn) + \sum_{i \ell m} S_i {\cal Y}_{i, \ell} Y_{\ell m} (\bn) ,
\ee
where $\tlen$ is the lensed CMB temperature, 
$Y_{\ell m}$ is a spherical harmonic, and 
$\prof_{i, \ell}$ is a profile function for each source, which describes the shape of the source on the sky if it is extended.
For a true ``point'' source which is a delta function in position space, $\prof_{i, \ell} = 1$.
We have assumed here that the sources are all radially symmetric for simplicity. 
We will also assume that only multipoles $L>100$ are used, so that we may ignore CMB temperature
anisotropies generated by the ISW effect.
Discarding multipoles at $L<100$ has negligible impact on the signal-to-noise ratio of the lensing estimator for 
experiments with the arcminute-scale sensitivity necessary to measure lensing.
Ignoring the ISW effect, $\tlen$ is linear in the primordial, unlensed CMB temperature and therefore
every non-zero $n$-point function must have an even number of multipoles associated with $\tlen$.
We may then group point source terms of the trispectrum by the number of individual sources which they contain.
Up to permutation symmetries of the trispectrum and source indices, there are 7 types of source term,
which we classify below depending on how many distinct sources they contain:
\begin{enumerate}
\item[1] source terms:
There are two types of source term containing a single source: 
$S_i^4$ and 
\mbox{$S_i^2 \tlen \tlen \equiv S_i^2 \phi$.} 
The $S_i^4$ term is essentially the kurtosis of the unresolved source population, and is analogous to the
``shot noise'' term in the power spectrum of the sources.
The 
$S_i^2 \phi$ term probes the correlation between the sources and the lensing potential.
\item[2] source terms:
There are three types of term containing two sources: $S_i^2 S_j^2$, $S_i^3 S_j$, and 
\mbox{$S_i S_j \tlen \tlen \equiv S_i S_j \phi$}.
The $S_i^2 S_j^2$ and $S_i^3 S_j$ terms probe the clustering of the sources, 
while the the $S_i S_j \tlen \tlen$
term probes the source-lensing bispectrum.
\item[3] source terms:
There is only one non-zero term containing three sources: $S_i S_j S_k^2$, which probes the bispectrum of the sources.
\item[4] source terms:
There is again only one non-zero term containing four sources: $S_i S_j S_k S_m$. 
This term probes the 4-point function of the sources, in which each of the four multipoles in the trispectrum
is sourced by a separate point.
\end{enumerate}

In all of the expressions it should be understood that the point source indices are disjoint
\mbox{(i.e. $i \ne j \ne k \ne m$)}, 
and the total trispectrum is obtained by summing over all indices. 
For several of the terms above, statistical isotropy implies that the details of the point source model
can enter only through associated power spectra.
For a population of sources all with identical profiles ${\cal Y}_{i, \ell}$ for example, we will have
\begin{align}
\left< \sum_i    \int d\bn \, Y_{LM}(\bn)  S_i^2(\bn) \phi_{L'M'}    \right> &= C_L^{S^2 \phi} \delta_{L L'} \delta_{M M'} \nonumber \\
\left< \sum_{ij} \int d\bn \, Y_{LM}(\bn)  S_i^2(\bn) (S_j^2)_{L'M'} \right> &= C_L^{S^2 S^2}  \delta_{L L'} \delta_{M M'} \nonumber \\
\left< \sum_{ij} \int d\bn \, Y_{LM}(\bn)  S_i^3(\bn) (S_j)_{L'M'}   \right> &= C_L^{S^3 S}    \delta_{L L'} \delta_{M M'}.
\label{eqn:sourcespectra}
\end{align}
However, for two of the terms the point source model enters through reduced bispectra $b_{\ell_1 \ell_2 L}$ of the
sources/lensing:
\begin{align}
\left< \sum_{ij}  (S_i)_{\ell_1 m_1} (S_j)_{\ell_2 m_2} \phi_{LM}    \right> &= {\cal G}_{\ell_1 \ell_2 L}^{m_1 m_2 M} b^{SS\phi}_{\ell_1 \ell_2 L} \\
\left< \sum_{ijk} (S_i)_{\ell_1 m_1} (S_j)_{\ell_2 m_2} (S_k^2)_{LM} \right> &= {\cal G}_{\ell_1 \ell_2 L}^{m_1 m_2 M} b^{SSS^2}_{\ell_1 \ell_2 L},
\end{align}
where the Gaunt integral is given by
\begin{multline}
{\cal G}_{\ell_1 \ell_2 L}^{m_1 m_2 M} \equiv 
%\int d\bn Y_{\ell_1 m_1} Y_{\ell_2 m_2} Y_{LM}.
\sqrt{ \frac{(2\ell_1 +1)(2\ell_2+1)(2L+1)}{4\pi} } 
\\ \times
\threej{\ell_1}{\ell_2}{L}{0}{0}{0} \threej{\ell_1}{\ell_2}{L}{m_1}{m_2}{M}.
\end{multline}

\begin{table}[tb]
\renewcommand\arraystretch{1.6}
\centering
\resizebox{\columnwidth}{!}{
\begin{tabular}{|c|c|}
\hline
Term & Primary Contraction $P^{\ell_1 \ell_2}_{\ell_3 \ell_4} (L)$ \\
\hline \hline
$\phi\phi$ & $C_L^{\phi\phi} W^{\phi}_{\ell_1 \ell_2 L} W^{\phi}_{\ell_3 \ell_4 L}$ \\ \hline
$S_i^4 $ &  $\frac{1}{3} \langle S^4 \rangle W_{\ell_1 \ell_2 L}^{S^2} W_{\ell_3 \ell_4 L}^{S^2}$ \\ \hline
$S_i^2 \phi$ & $\frac{1}{2} C_L^{S^2 \phi} (W_{\ell_1 \ell_2 L}^{S^2} W_{\ell_3 \ell_4 L}^{\phi} + W_{\ell_1 \ell_2 L}^{\phi} W_{\ell_3 \ell_4 L}^{S^2})$ \\ \hline
$S_i^2 S_j^2$ & $C_L^{S^2 S^2}  W_{\ell_1 \ell_2 L}^{S^2} W_{\ell_3 \ell_4 L}^{S^2}$ \\ \hline
$S_i^3 S_j$ & $\frac{1}{3} (C_{\ell_1}^{S^3 S} + C_{\ell_2}^{S^3 S} + C_{\ell_3}^{S^3 S} + C_{\ell_4}^{S^3 S})  W_{\ell_1 \ell_2 L}^{S^2} W_{\ell_3 \ell_4 L}^{S^2}$ \\ \hline
$S_i S_j \phi$ & $\frac{1}{2}( b^{SS\phi}_{\ell_1 \ell_2 L} W_{\ell_1 \ell_2 L}^{S^2} W_{\ell_3 \ell_4 L}^{\phi} + b^{SS\phi}_{\ell_3 \ell_4 L} W_{\ell_1 \ell_2 L}^{\phi} W_{\ell_3 \ell_4 L}^{S^2} ) $ \\ \hline
$S_i S_j S_k^2$ & $\frac{1}{2}( b^{SSS^2}_{\ell_1 \ell_2 L} + b^{SSS^2}_{\ell_3 \ell_4 L} ) W_{\ell_1 \ell_2 L}^{S^2} W_{\ell_3 \ell_4 L}^{S^2}$ \\ \hline
$S_i S_j S_k S_m$ & $\frac{1}{3} {}^s T^{\ell_1 \ell_2}_{\ell_3 \ell_4} (L) $ \\ \hline
\end{tabular}
}
\caption{Primary source terms discussed in Sect.~\ref{sec:source_trispectrum}. 
The first row gives the trispectrum due to lensing for comparison purposes.
For simplicity of presentation we have assumed that all of the sources are delta functions, such
that we may drop indices on the weight functions $W^{S^2}_{\ell \ell' L}$.}
\label{tab:source_terms}
\end{table}
We present the primary contractions of the trispectra for the seven different source terms in Table~\ref{tab:source_terms}. 
To simplify our presentation, we find it useful to introduce the ``point source weight function'' given by
\begin{multline}
W_{\ell_1 \ell_2 L}^{S_i S_j} = 
 \sqrt{\frac{(2\ell_1+1)(2\ell_2+1)(2L+1)}{4\pi}}
 \\ \times
\threej{\ell_1}{\ell_2}{L}{0}{0}{0} \prof_{i, \ell_1} \prof_{j, \ell_2}.
\label{eqn:qe_weight_ptsrc}
\end{multline}
Often in this work we will assume delta function point sources and drop the subscript $i$ for the weight function, 
denoting it in this case as 
$W_{\ell_1 \ell_2 L}^{S^2}$.
Most of the trispectrum terms in Table~\ref{tab:source_terms} include one or more factors of this weight function. 
This will prove useful in the next section, when we discuss estimators for the source trispectra.

\section{Source Estimators}
\label{sec:triest}

One approach to mitigating point source biases in the lensing spectrum is to construct
physical models for the various source trispectra presented in the previous section,
and then to propagate these to biases in the lens reconstruction,
which may then be subtracted to obtain unbiased estimates of $C_\ell^{\phi\phi}$.
However, this is subject to uncertainty in modelling the source populations, as well as
issues with precisely determining, for example, the flux density cut that at which sources are reliably detected and masked.

A complementary and potentially more robust approach is to jointly estimate both the lensing and point-source trispectra,
or alternatively to construct lensing estimators which probe trispectrum configurations orthogonal to
those generated by point sources.

To elaborate on this approach, we consider first constructing a quadratic ``point source estimator''
$\bar{S}^2_{LM}$ following Eq.~\eqref{eqn:qe}, with weight function $W^{S^2}_{\ell_1 \ell_2 L}$. 
As one might expect, in real space this estimator corresponds to squaring the inverse-variance filtered sky map
\begin{align}
\hat{S}^2_{LM} 
&=
( {\cal R}^{S^2 S^2}_L )^{-1}
\bar{S}^2_{LM} \nonumber \\
%&= \frac{1}{2} \sum_{\ell_1 m_1, \ell_2 m_2} \threej{\ell_1}{\ell_2}{L}{m_1}{m_2}{-M} W_{\ell_1 \ell_2 L}^s \bar{T}_{\ell_1 m_1} \bar{T}_{\ell_2 m_2} \nonumber \\
&= ( {\cal R}^{S^2 S^2}_L )^{-1} \frac{1}{2} \int d\hn \, Y_{LM}^*(\hn) \, \bar{T}^2(\hn).
\label{eqn:S2_est}
\end{align}
Following the intuitive picture outlined in the introduction, this estimator looks for variations in the map 
\mbox{``noise level''} which can be attributed to point sources.
For unclustered sources, the measured noise level will simply vary across the map with a variance which is larger-than-expected given the instrumental noise level.
For clustered sources, the measured noise level will again show excess variance, and these variance fluctuations will in turn show correlations on the clustering scale.

Estimates for the source terms such as $C_L^{S^2 S^2}$ and $C_L^{S^2 \phi}$ may be formed intuitively by taking the auto-spectrum
of $\hat{S}^2_{LM}$, or its cross-spectrum with the lensing potential estimate $\hat{\phi}$.
The resulting estimates may then be used to calculate biases for $\hat{C}_L^{\phi\phi}$.
Another approach is simply to construct an estimator which is less sensitive to the source contributions.
If we think of the sky as containing statistical anisotropy sourced by lensing ($\phi$) and point sources ($S^2$), both
estimators pick up unwanted contributions:
\be
\label{eqn:response}
\begin{split}
\langle \bar{\phi}_{LM} \rangle &= {\cal R}^{\phi \phi}_L {\phi}_{LM} + {\cal R}^{\phi S^2}_L {S}^2_{LM},  \\
\langle \bar{S}^2_{LM} \rangle  &= {\cal R}^{S^2 S^2}_L   {S}^2_{LM}  + {\cal R}^{S^2 \phi}_L {\phi}_{LM}. \\
\end{split}
\ee
Following \cite{Namikawa:2012pe}, we can then construct ``bias-hardened'' (BH) lensing estimators for both $\phi$ and $S^2$ as
\be
\label{eqn:bre_matrix}
\left[ \begin{matrix} \hat{\phi}^{BH}_{LM} \\ \hat{S}^{2, BH}_{LM} \end{matrix} \right] =
\left[ \begin{matrix} \mathcal{R}^{\phi\phi}_L & \mathcal{R}^{\phi S^2}_L \\ \mathcal{R}^{S^2 \phi}_L & \mathcal{R}^{S^2 S^2}_L \end{matrix} \right]^{-1}
\left[ \begin{matrix} \bar{\phi}_{LM} \\ \bar{S^2}_{LM} \end{matrix} \right].
\ee
The weight function for the bias-hardened $\phi$ estimator is given by
\be
W^{\mre}_{\ell_1 \ell_2 L} = W^{\phi}_{\ell_1 \ell_2 L} - {\cal R}_L^{\phi S^2} (  {\cal R}_L^{S^2 S^2} )^{-1} W^{S^2}_{\ell_1 \ell_2 L}.
\label{eqn:est_attempt1}
\ee
We will refer to the estimator using this weight function as the ``source-hardened'' lensing estimator -- it has the property that ${\cal R}^{\mre S^2}_L$ is zero.
Note that here we have assumed delta function point sources, although we could also construct source-hardened estimators which are orthogonalized against a set of finite source profiles ${\cal Y}_{i, \ell}$, by repeated application of the bias-hardening procedure \cite{Namikawa:2012pe}.

Proceeding to the trispectrum, there are a few subtleties due to the secondary contractions of Eq.~\eqref{eqn:trispectrumencoding}.
To formalize this discussion, it is useful to generalize the estimator for the lensing power
spectrum of Eq.~\eqref{eq:clpp_est}.
Consider a trispectrum with the form
\be
{}^{ab}\! P^{\ell_1 \ell_2}_{\ell_3 \ell_4} (L) = 
\frac{1}{2} C_L^{ab} (W^{a}_{\ell_1 \ell_2 L} W^{b}_{\ell_3 \ell_4 L} + W^{b}_{\ell_1 \ell_2 L} W^{a}_{\ell_3 \ell_4 L}),
\ee
where $(a,b)$ denote a pair of weight functions. The first five primary contractions in Table~\ref{tab:source_terms}
except (possibly) for the final trispectrum term can be written in this form,  or as a sum over a small number of
terms with this form. 
If the source bispectra may be written as a sum of separable terms, then they can also be included in this discussion.  
We denote the response of the cross-spectrum between a pair of estimators $(x,z)$ to the $(a,b)$ trispectrum as
\be
\tcl^{xz}_{L, ab} = 
\frac{2}{ {\cal R}^{xa}_L {\cal R}^{zb}_L + {\cal R}^{xb}_L {\cal R}^{za}_L } 
\left[ \frac{1}{2L+1} \sum_{M} \bar{x}_{LM} \bar{z}_{LM}^* - N_{L}^{xz} \right].
\label{eq:tcl}
\ee
We refer to $\tcl^{xz}_{L, ab}$ as the trispectrum-related spectrum, 
with the lensing result of Eq.~\eqref{eq:clpp_est} as a specific case.
Eq.~\eqref{eq:tcl} is an estimator for the $x$-$z$ cross-spectrum obtained from estimators $x$ and $z$.
The estimators $x$ and $z$ are arbitrary, and may or may not be bias-hardened.
In the case that $(x,z) = (a,b)$ we will use the shorthand $\tcl^{ab}_L$.
If the data contains non-Gaussianity with an $(a,b)$ trispectrum, then
averaging over CMB temperature realizations with fixed $a$ and $b$ we have
\be
\langle \tcl^{xz}_{L, ab} \rangle = C_L^{ab} + M^{L'}_{L} C_{L'}^{ab},
\ee
where the mixing matrix $M$ depends on the estimators and trispectrum being considered, and as before $L'$ is summed over.
If the data contains non-Gaussianity of some other type $(c,d)$,
then $\tcl$ may be biased.
This bias is given schematically by 
\be
\left. \tcl^{xz}_{L, ab} \right|_{cd} = 
%\left. {R}^{xz}_{L, ab} \right|_{cd}
\frac{ 
	{\cal R}^{xc}_L {\cal R}^{zd}_L + {\cal R}^{xd}_L {\cal R}^{zc}_L 
	}{ 
	{\cal R}^{xa}_L {\cal R}^{zb}_L + {\cal R}^{xb}_L {\cal R}^{za}_L 
	} 
	C_L^{cd} + M^{L'}_{L} C_{L'}^{cd}.
\label{eqn:clxzabcd}
\ee
Again, the matrix $M$ depends implicitly on all of the indices being considered: 
the estimators $(x, z)$, 
the trispectrum being estimated $(a,b)$,
and the contaminating trispectrum $(c, d)$.

The first term on the RHS of 
Eq.~\eqref{eqn:clxzabcd} 
represents the response of the 
$\tcl^{xz}_{L, ab}$ spectrum to the primary contraction of the $(c,d)$ trispectrum.
It can be seen that if a bias-hardened estimator is used for $(x,z)$, such that
the response functions in the numerator of Eq.~\eqref{eqn:clxzabcd} are zero, then this term will be zero as well.
This is the case, for example, with the source-hardened lensing estimator discussed above.
It can be seen that it receives no contributions from the primary contractions of the 
$S_i^4$, $S_i^2 \phi$, and $S_i^2 S_j^2$ 
terms in Table.~\ref{tab:source_terms} (modulo issues with the matching of source profiles).

The second term in Eq.~\eqref{eqn:clxzabcd} represents the response to the secondary contractions of the $(c,d)$ trispectrum. 
Depending on the $(x,z)$ and $(c,d)$ weight functions, the secondary contributions to this bias can be
costly to evaluate using the harmonic space expressions, however we can evaluate them using 
the flat-sky expressions presented in Appendix~\ref{sec:secondary_contractions}.

A special case of Eq.~\eqref{eqn:clxzabcd} occurs when the contaminating trispectrum is due to shot noise 
(the $S^4_i$ terms in Table.~\ref{tab:source_terms}).
In this case, the biases due to each of the two secondary contractions are equal to that of the primary contraction.
For a collection of point sources with flux $S_i$ and profiles $\prof_{i, \ell}$ the resulting bias becomes
\be
\left. \tcl^{xz}_{L, ab} \right|_{S^4} = 
\sum_i
\frac{ 
	{\cal R}^{x S^2_i}_L {\cal R}^{z S^2_i}_L + {\cal R}^{x S^2_i}_L {\cal R}^{z S^2_i}_L 
	}{ 
	{\cal R}^{xa}_L {\cal R}^{zb}_L + {\cal R}^{xb}_L {\cal R}^{za}_L 
	} S^4_i.
\label{eqn:clxzabcd_shotnoise}
\ee

\section{Simulation Results}
\label{sec:fore_sim}
In the previous sections, we have discussed generic shapes for point source trispectra,
estimators to probe them in data, 
as well as the construction of source-hardened lensing estimators.
In this section, we will test these ideas using the simulations of \citet{2010ApJ...709..920S} at $148\,$GHz, which is
the frequency at which most recent lensing analyses have been performed. 
This simulation set contains  full-sky HEALpix maps of several components, including 
radio point sources (RPS), 
galaxy clusters observable through the thermal Sunyaev-Zel'dovich (tSZ) effect,
and infrared sources (IR), which form the Cosmic Infrared Background (CIB).
There are also simulated maps of the kinetic Sunyaev-Zel'dovich effect, 
although the amplitude of lensing contamination from this component 
is much smaller than the others and so we do not present results for it here.
We analyze the simulations assuming a noiseless experiment, band-limited to either 
$l_{\rm max}=1500$ (to mimic the beam cutoff of the {\it Planck} satellite) or
$l_{\rm max}=3000$ (to mimic the higher resolution analyses of ACT and SPT).
In order to mimic the source masking which is used when analyzing data from these experiments,
we mask all RPS and IR sources with flux greater than $200$mJy for $l_{\rm max}=1500$ and $5$mJy for $l_{\rm max}=3000$.
We also construct an SZ cluster mask by degrading the tSZ component map to a lower HEALPix resolution $N_{\rm side}^{\rm low}$
and then mask all low-resolution pixels in which the absolute value of the tSZ temperature fluctuation is greater than $5N$, 
where $N$ is a given noise RMS per pixel.
For this procedure we take $N_{\rm side}^{\rm low} = 512$ and N corresponding to a map noise level of $70\muKarcmin$ for $l_{\rm max} = 1500$ and
$N_{\rm side}^{\rm low} = 1024$ with a map noise level of $15\muKarcmin$ for the $l_{\rm max}=3000$ experiment.
In order to avoid issues on large-scales from the quasi-periodic nature of the simulations, 
we create an apodized mask which removes all but one octant of the sky, 
and correct our lensing estimates with appropriate $f_{\rm sky}$ factors following \cite{BenoitLevy13}. 
We also ignore all multipoles with $L < 300$. 
We estimate disconnected noise biases from the power spectra of the masked, apodized sky. 
To inverse-variance filter the simulations (as in Eq.~\ref{eqn:tbar}) we use $F_{\ell} = [ C_{\ell}^{TT} ]^{-1}$, 
ignoring the power due to point sources in the filter function. 
This choice makes our lensing estimator slightly more susceptible to foreground contamination than it would be with optimal filtering to the angular scales which are contaminated by point sources.

\subsection{Source Trispectra}
\begin{figure*}[ht!]
\vspace{0.2in}
  \centering
  \includegraphics[width=\textwidth]{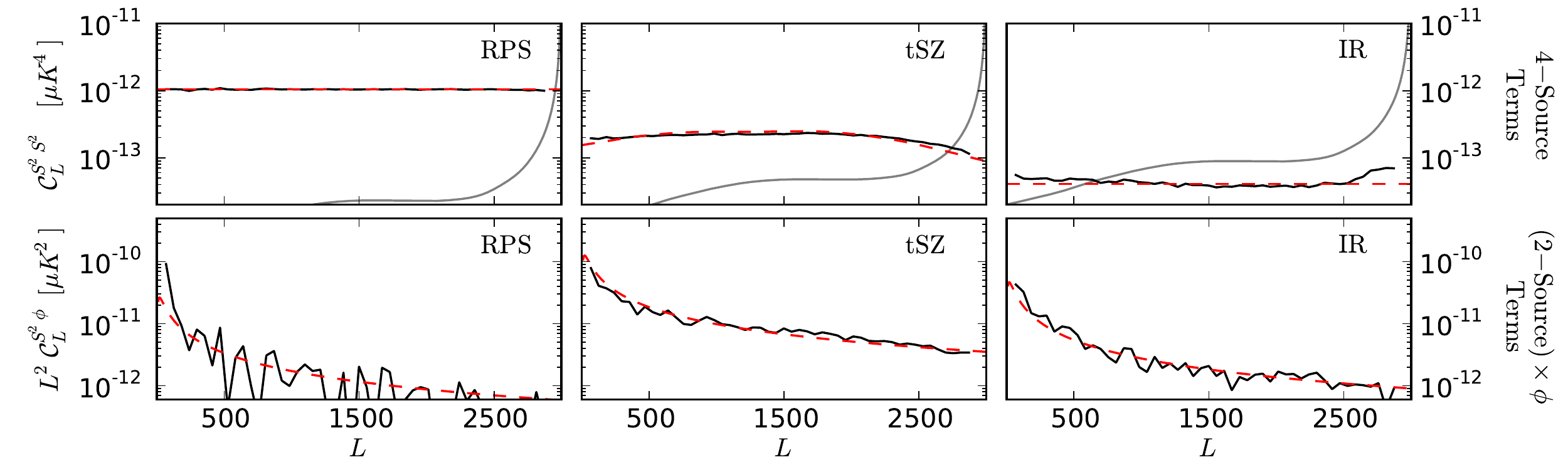}
  \caption{
  Simulation-based estimates of the point source trispectra for 
  radio point sources (RPS), 
  the thermal Sunyaev-Zel'dovich effect (tSZ), and
  infrared sources (IR). 
  The estimation has been performed for the $l_{\rm max}=1500$ experiment, results for $l_{\rm max}=3000$ are qualitatively similar.
  Absolute values have been taken for each plotted curve. 
  The top panels show auto-spectra of the point source estimator (Eq.~\ref{eqn:S2_est}), after subtracting the disconnected term $N_L^{S^2 S^2}$ (black solid).
  This probes the connected biases when all four trispectrum legs are due to point sources. 
  The $N_L^{S^2 S^2}$ terms which have been subtracted are plotted in gray. 
  The lower panels show the cross-spectra of the point source estimator with the input lensing potential realization $\phi$. 
  This probes the connected biases when only two of the four trispectrum legs are due to point sources.
  Dashed red curves are model fits, described in the text.
}
  \label{fig:clss_lmax_1500}
\end{figure*}
We start our analysis by looking at estimators for some of the point source trispectra.
Following Table.~\ref{tab:source_terms}, 
the source trispectra can be separated into two qualitative groups: 
those in which all four legs are due to point sources, and those in which two of the legs are from lensed CMB temperature fluctuations (which in turn trace $\phi$).
We therefore study these two sets of terms separately.
We probe both sets of terms using the quadratic point source estimator $\hat{S}^{2}_{LM}$  of Eq.~\eqref{eqn:S2_est}.

To look at the ``4-source'' terms, we evaluate the trispectrum estimator $\tcl^{S^2 S^2}_{L}$
(in the notation of Eq.~\ref{eq:tcl}), assuming delta function point sources.
This is accomplished simply by taking the auto-spectrum $\hat{S}^2_{LM}$ and subtracting an estimate of the disconnected noise bias.
The results of this calculation are plotted for the 
\mbox{$l_{\rm max}=1500$}
 experiment in the upper panels of Fig.~\ref{fig:clss_lmax_1500}. 
 The results for the $l_{\rm max}=3000$ experiment are qualitatively similar. 
We can see that the RPS contribution is nearly flat, corresponding to the shot noise trispectrum ($S^4$) of Table.~\ref{tab:source_terms}.
The RPS population is characterized by steep number counts $dN/dS$, and its trispectrum is generated by a handful of sources just below the flux cut. 
As a simple illustration, using a power law form for the number counts 
\be
\frac{dN}{dS} = \frac{N_0}{S^{\beta}}
\ee
with parameters $N_0 = 12{\rm Jy}^{1.15} {\rm sr}^{-1}$ and $\beta = 2.15$, 
given in \cite{2007MNRAS.379.1442W,2008arXiv0811.3916S}
we can analytically calculate
\be
\langle S^4 \rangle = \sum_{S=0}^{S_{\rm max}} dS \frac{dN}{dS} S^{4}, 
\ee
Converting to CMB temperature units we find 
$\langle S^{4} \rangle = 2 \times 10^{-12} \mu K^4$ at 148\,GHz
for a flux cut of $200\,$mJy and a much smaller level of 
$5 \times 10^{-17} \mu K^2$ for a cut of $5\,$mJy.
Both of these numbers are consistent (to within a factor of two) with the amplitudes we see for the simulations.

We find that the tSZ population is also well described as shot noise dominated, although it is important to incorporate information about the distribution of source profiles to reproduce the measured source-related power spectrum in Fig.~\ref{fig:clss_lmax_1500}. 
Using the tSZ model of \citet{2002MNRAS.336.1256K} 
(in which halos at a given virial mass and redshift are associated with a flux-weighted profile 
${\cal Y}_l (M, z)$)\footnote{
In the notation of \cite{2002MNRAS.336.1256K}, these profiles are denoted with a lowercase $y$, as in their Eq.~(2).}, 
we can evaluate Eq.~\eqref{eqn:clxzabcd_shotnoise} by writing the sum over sources as a mass and redshift integral with the substitution
\be
\sum_i \rightarrow g_{\nu}^{4} \int_{M=M_{\rm min}}^{M_{\rm max}} dM \int_{z=0}^{z_{\rm max}} dz \frac{dV}{dz} \frac{dN(M,z)}{dM},
\ee
where $g_{\nu}$ is the spectral response of the tSZ effect, 
$V(z)$ gives the volume of the Universe per steradian at redshift $z$ and $dN/dM$ is the halo mass function.
These quantities are discussed in more detail around Eq.~(1) of \cite{2002MNRAS.336.1256K}.
We use the mass function of \cite{Tinker:2010my}, 
with
$M_{\rm min} = 5 \times 10^{11} \Msolar$.
For the simple tSZ masks which we have constructed 
we find that the measured trispectra are well fit with
$M_{\rm max} = 1.2 \times 10^{15} \Msolar$ for the 
$l_{\rm max}=1500$ experiment and 
$M_{\rm max} = 4.7 \times 10^{14} \Msolar$ 
for the 
$l_{\rm max}=3000$ 
experiment.
The resulting bias, 
${\tcl}^{S^2 S^2}_L |_{S^4}$,  
is plotted as the red dashed curve in the upper tSZ panel of Fig.~\ref{fig:clss_lmax_1500}.

The IR 4-source terms are plotted in the upper right panel of Fig.~\ref{fig:clss_lmax_1500}. 
This source population is the most Gaussian of the three studied here, with the disconnected term
being larger than the connected trispectrum on angular scales above $L=500$. 
The connected trispectrum of the IR population does not appear to be as simple to model as the RPS and tSZ terms however.
As a guide to the eye, we have plotted a best-fit shot noise amplitude with red dashes.
On large angular scales there is a slight increase of power in the measured trispectrum, 
which could be characteristic of clustering, 
although as we will see in 
Sec.~\ref{sec:results:biases} 
it does not appear to be of the $C_l^{S^2 S^2}$ form in 
Table.~\ref{tab:source_terms}.
On very small scales, there is again an increase in the trispectrum amplitude, 
which we have so far been unable to find a modelling explanation for.

Turning now to the ``2-source'' terms, we take the cross-spectrum of the point source estimator with the input $\phi$ realization. 
Unlike the 4-source term, there is no disconnected bias to remove in this case. 
The results are plotted as the black solid curves in the lower panels of Fig.~\ref{fig:clpp_lmax_3000}. 
This procedure should be most sensitive to the $C_L^{S^2 \phi}$ trispectrum shape of Table.~\ref{tab:source_terms}.
We find that for all three source populations, the measured cross-spectrum is reasonably approximated as a power law
with 
\be
C_L^{S^2 \phi} \propto (L + 10)^{-3}.
\label{eqn:sl2ptemplate}
\ee
We have fitted amplitudes for this template to the measured cross-spectra, which are plotted in red dashes. 
To investigate the origins of this power law behaviour in more detail, 
we have evaluated $C_L^{S^2 \phi}$ analytically under the Limber approximation using a linear matter power spectrum and 
a grid of Gaussian redshift distributions which, with means in the range 
$0.1 \le \langle z \rangle \le 4.$ and standard deviations in the range $0.1 \le \sigma_z \le 2$.
We find that all exhibit a power law behaviour on small scales ($L>1000$), with a 
slope of $\approx -4$, which is somewhat deeper than that in Eq.~\eqref{eqn:sl2ptemplate}. 
We take this to indicate that on these scales, non-linearity has a significant contribution to the 
$C_L^{S^2 \phi}$ power.
\begin{figure}[!tbh]
%\vspace{0.05in}
  \centering
  \includegraphics[width=0.49\textwidth]{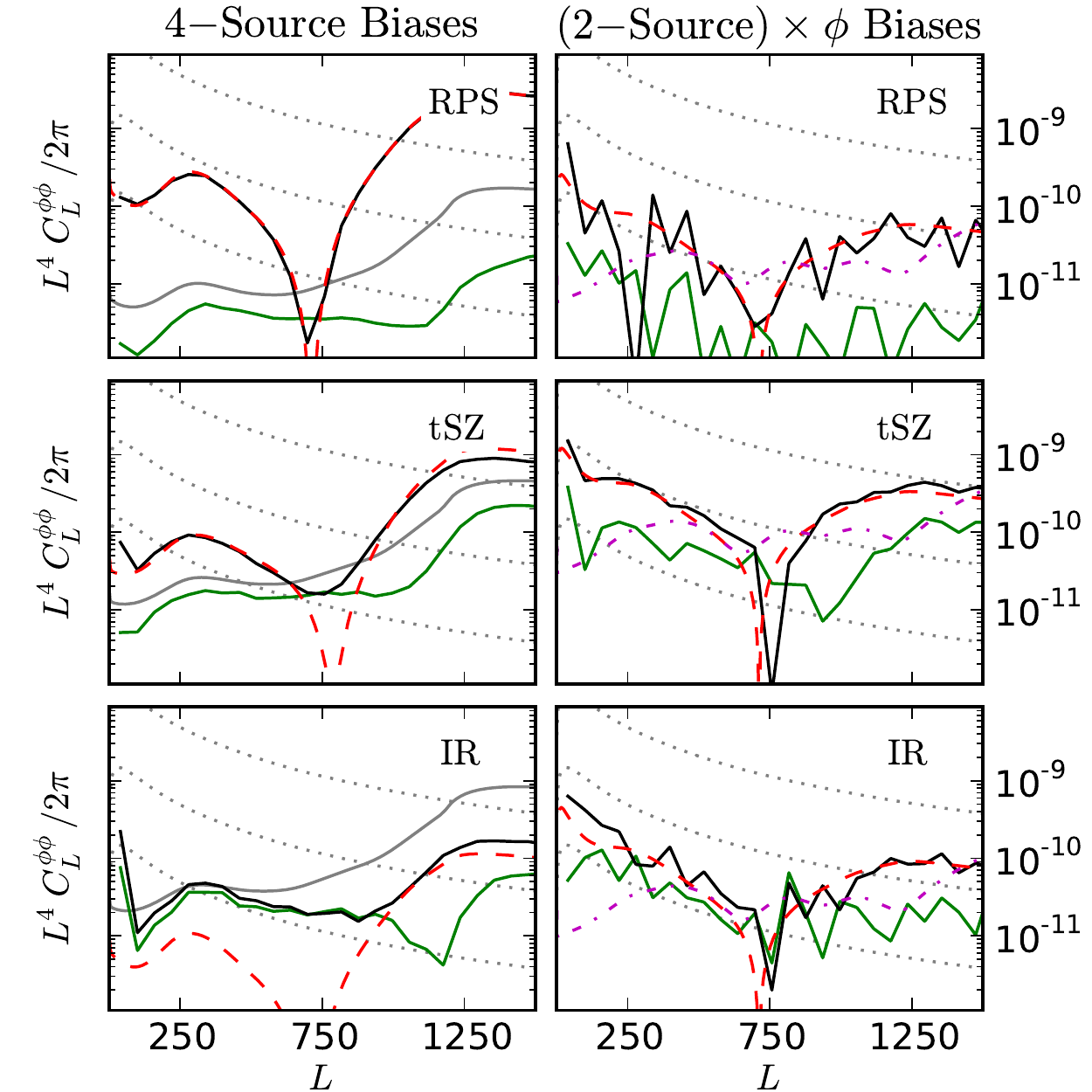}
  \caption{Simulation-based lensing bias estimates for 
  radio point sources (RPS), 
  the thermal Sunyaev-Zel'dovich effect (tSZ), 
  and infrared sources (IR), for the $l_{\rm max}=1500$ experiment. 
  Absolute values have been taken for each plotted curve. 
  Black lines give results for the standard lensing estimator, while green lines give results for the source-hardened estimator of Eq.~\eqref{eqn:est_attempt1}.
   In most cases, the contamination is significantly smaller for the source-hardened estimator. 
  Dashed red curves are model fits, obtained by analytically propagating the corresponding model curves in Fig.~\ref{fig:clss_lmax_1500} to lensing. 
  Dotted gray lines give fractional biases (from top to bottom) of $C_L^{\phi\phi} \times (0.1, 0.01, 0.001)$.
  Left-hand panels show the auto-spectra of the quadratic estimators applied to each component map, after subtracting the disconnected term $N_L^{\phi\phi}$.
  The $N_L^{\phi\phi}$ terms which have been removed are plotted in gray.
  Right-hand panels show the cross-spectra of the quadratic estimators with the input lensing potential (multiplied by a factor of $2\times$ to represent the bias to the auto-spectrum).
  Dot-dashed magenta is an estimate for the bias due to secondary contractions, discussed in the text. }
  \label{fig:clpp_lmax_1500}
\end{figure}

\subsection{Lensing Biases}
\label{sec:results:biases}
\begin{figure}[!tb]
%\vspace{0.2in}
  \centering
  \includegraphics[width=\columnwidth]{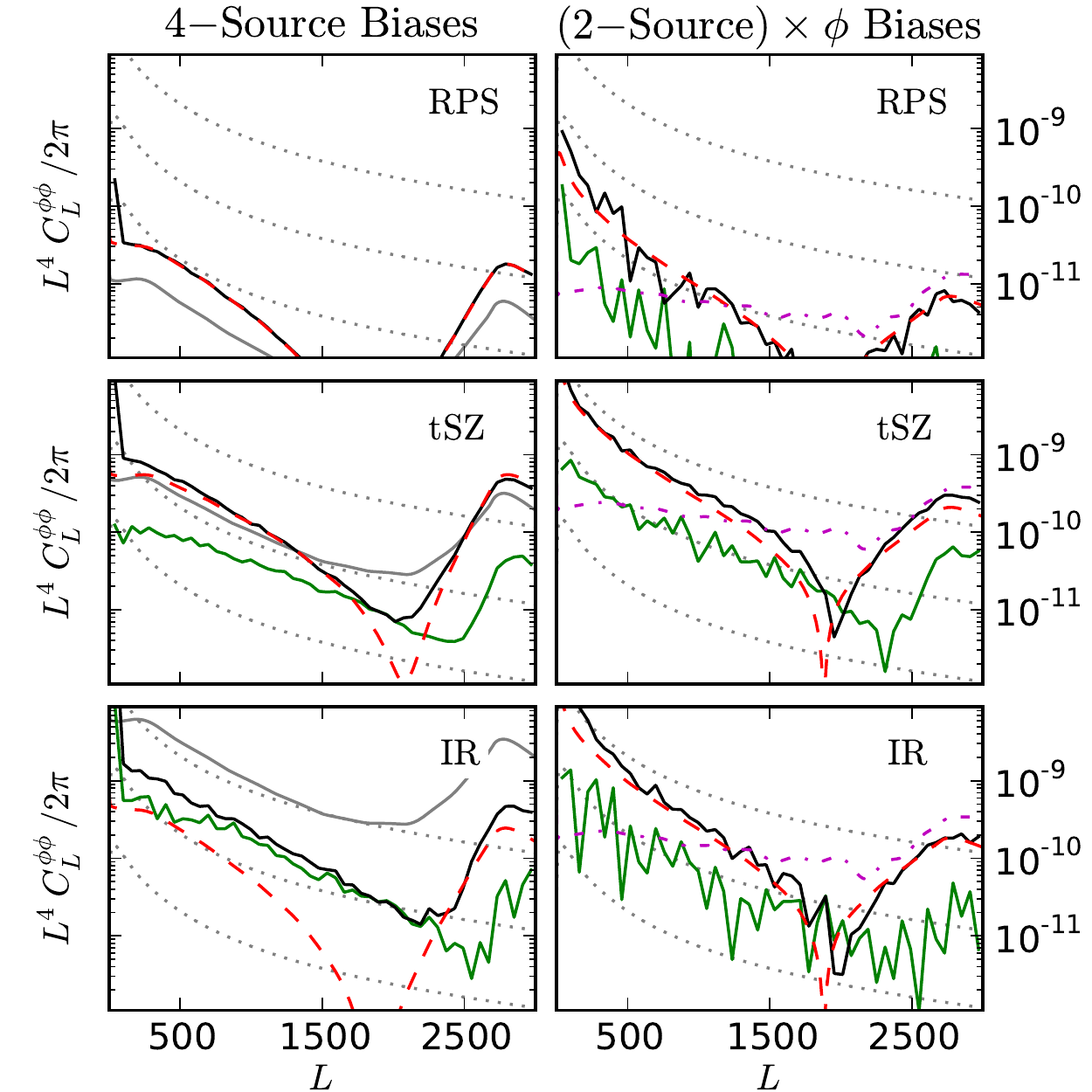}
  \caption[]{Same as Fig.~\ref{fig:clpp_lmax_1500}, but evaluated for an $l_{\rm max}=3000$ lens reconstruction. }
  \label{fig:clpp_lmax_3000}
\end{figure}

Having examined the source related trispectra, 
we now turn to the biases which they generate 
for the estimates of the CMB lensing potential power spectrum. 
We feed the RPS, tSZ and IR component maps into both the 
standard quadratic lensing estimator $\hat{\phi}$ as well as the 
source-hardened estimator of Eq.~\eqref{eqn:est_attempt1}. 
As in the previous section, 
we take both the auto-spectrum of these estimates (to probe the 4-source terms),
as well as the cross-spectrum with the input lensing potential (to probe the 2-source terms).
In this case, because we are interested in the bias on the lensing power spectrum, 
we multiply the 2-source terms by a factor of two.
Our results are shown in Fig.~\ref{fig:clpp_lmax_1500} and Fig.~\ref{fig:clpp_lmax_3000} for the two values of $l_{\rm max}$.  
In principle these estimators are able to probe the lensing potential up to $L=3000$ and $L=6000$ respectively, however 
in practice these modes are often very noisy and so we have only plotted them to $L=1500$ and $L=3000$ to emphasize the modes
which are of interest in the lens reconstruction.
It can be seen that the biases for the standard lensing estimator (plotted in black solid)
are generally larger than those for the source-hardened estimator (plotted in green solid).
In particular, for radio point source population, which are delta functions on the sky and completely dominated by the ``shot noise'' term $S_i^{4}$, bias hardening reduces the measured bias by more than two orders of magnitude for the $4$-source terms and one order of magnitude for the $2$-source terms. 

The improvement for tSZ sources is somewhat smaller, generally 1 order of magnitude on large scales. 
We note that the source-hardened estimator we have used here is configured to reject delta function point sources. 
Better results would likely be obtained constructing a more involved estimator which is hardened against an ensemble of source profiles characteristic of the tSZ, however given that the bias reduction is already sizeable we have not attempted this improvement.

The improvement for the IR source population is not as dramatic, although the $S^2 S^2$ contribution is still reduced on small angular scales, and the $S^2 \phi$ contribution is modestly reduced on all scales. 
This is consistent with the results of the previous section, where we had difficulty finding an explanation for the shape of the IR trispectrum. 

We note one final aspect of the 2-source bias terms.  
When cross-correlating the lensing estimate obtained on source maps with the input $\phi$ realization we are only probing one contraction of the 2-source trispectra. 
By convention we will label this as the primary contraction. 
To get a feeling for the magnitude of the secondary contribution, 
we calculate the secondary contractions associated with the $S^2 \phi$ trispectrum, for the template shape given in 
Eq.~\eqref{eqn:sl2ptemplate}.
These are plotted as the dot-dashed magenta curves in Figs.~\ref{fig:clpp_lmax_1500}/\ref{fig:clpp_lmax_3000}.
We have verified using the simulations 
(by taking the auto-spectra of quadratic estimates in which one leg is from a component map, and one uses the lensed CMB temperature)
that these estimates of the secondary contribution are reasonable (although this procedure is somewhat noisy, due to the CMB modes which act as a source of noise). 
On large scales, the primary contractions generally dominate, 
however on small scales the secondary terms are often as large or larger in magnitude than the primary ones -- 
similar to the case with the CMB lensing trispectrum, where the secondary contractions are known as the ``$N^{(1)}$'' bias \cite{2003PhRvD..67l3507K}.

\section{Conclusions}
\label{sec:conc}
At the resolution of current CMB measurements from ACT, SPT and {\it Planck}, foregrounds are believed to generate percent level biases for lens reconstruction.
We have considered a data-based approach to mitigate this contamination, 
by constructing estimators for the source contribution which are used to project out the contaminating trispectra.
This provides another tool, in conjunction with frequency dependence and modelling, to quantify or reduce biases from the non-Gaussianity of unresolved foreground point sources. 
This approach works well for radio point sources and tSZ clusters, particularly on large scales, and generally reduces the contamination by an order of magnitude or more.
The only trispectrum which we have had trouble interpreting, and which does not improve significantly with our default source-hardened estimator is the IR population, in particular the $4$-source contribution. 
We note, however, that with high-frequency measurements from {\it Planck} and {\it Herschel}, maps of the cosmic infrared background fluctuations which source the IR component are now available over much of the sky and may be used to strongly suppress this contribution in practice, as was done in \cite{Ade:2013mta}.

Finally, we also note that although we have focussed here on biases to the lensing potential power spectrum, the methodology we have discussed can also be carried over directly to other estimators of trispectrum non-Gaussianity. 

\section{Acknowledgements}
SJO acknowledges support from the US \Planck\ Project, which is funded by the NASA Science Mission Directorate.
Part of the research described in this paper was carried out at the Jet Propulsion Laboratory, California Institute of Technology, under a contract with the National Aeronautics  and Space Administration.
Some of the results in this paper have been derived using the HEALPix~\citep{2005ApJ...622..759G} package.

\bibliography{lensing_foregrounds}

\renewcommand{\theequation}{A-\arabic{equation}}
\setcounter{equation}{0}

\appendix

\section{Secondary Contractions}
\label{sec:secondary_contractions}

On the full-sky, calculations involving secondary contractions of Eq.~\eqref{eqn:trispectrumencoding} 
often involve computationally expensive Wigner-$6j$ symbols.
In this section, we present alternative flat-sky expressions which are numerically more tractable.

A flat-sky quadratic estimator, $\bar{x}_{LM}$ is given by 
\be
\bar{x}( \bL ) = \int \frac{d^2 \bl_1}{(2\pi)^2} W^{x}(\bl_1, \bl_2) \bar{T}(\bl_1) \bar{T}(\bl_2),
\ee
where $\bl_2 \equiv \bL - \bl_1$.
The weight functions $W^{x}(\bl_1, \bl_2)$ are the flat-sky analogues to the $W^{x}_{\ell_1 \ell_2 L}$ weight functions.
For lensing, the weight function is given by
\be
W^{\phi}(\bl_1, \bl_2) = C_{ \ell_1 }^{TT} (\bl_1 + \bl_2) \cdot \bl_1 + C_{ \ell_2 }^{TT} (\bl_1 + \bl_2) \cdot \bl_2
\ee
and for point sources, the weight function is
\be
W^{S^2}(\bl_1, \bl_2) = 1.
\ee
The flat-sky weight function for the source-hardened estimator is given by replacing the full-sky weight functions in 
Eq.~\eqref{eqn:est_attempt1} with those above.
There are flat-sky expressions for the response functions ${\cal R}_L^{xy}$, however in this work we simply use the full-sky ones.

The connected trispectrum on the flat-sky is denoted as
\begin{multline}
\langle T(\bl_1) T(\bl_2) T(\bl_3) T(\bl_4) \rangle_C = \\ (2\pi)^{-2} \, \delta( \bl_1 + \bl_2 + \bl_3 + \bl_4 ) \, T(\bl_1, \bl_2, \bl_3, \bl_4).
\end{multline}
As on the full-sky, it is useful to break the trispectrum into primary and secondary contractions as
\begin{multline}
T(\bl_1, \bl_2, \bl_3, \bl_4) = 
 \int d^2 \bL \, \delta( \bl_1 + \bl_2 + \bL) \delta( \bl_3 + \bl_4 - \bL ) 
\\ \times \left(
 P^{\bl_1 \bl_2}_{\bl_3 \bl_4} (\bL) + 
 P^{\bl_1 \bl_3}_{\bl_2 \bl_4} (\bL) + 
 P^{\bl_1 \bl_4}_{\bl_3 \bl_2} (\bL) 
 \right)
\end{multline}

The forms of the trispectra in Table~\ref{tab:source_terms} map directly onto the flat-sky expressions, simply by
replacing the full-sky weight functions with the corresponding flat ones. If our data contains a $(c,d)$ trispectrum with the form 
\be
{}^{cd}\! P^{\ell_1 \ell_2}_{\ell_3 \ell_4} (\bL) = 
C_L^{cd} W^{c}(\bl_1, \bl_2) W^{d}( \bl_3, \bl_4 ),
\ee
then the bias to the $(x,z)$ estimator for the $(a,b)$ trispectrum (denoted by the quantity $\tcl^{xz}_{L, ab}$ defined
in Eq.~\ref{eq:tcl}) is given by
\begin{multline}
\left. \tcl^{xz}_{L, ab} \right|_{cd} = 
\left. {R}^{xz}_{L,ab} \right|_{cd} C_L^{cd} + 
\left. {R}^{xz}_{L,ab} \right|_{cd}
\int \frac{d^2 \ell_1}{(2\pi)^2} 
\int \frac{d^2 \ell_2}{(2\pi)^2} 
\\ \times
\Big\{
C_{| \bl_1 - \bl_1' |}^{cd}  
W^{c}( -\bl_1,  \bl_1' )
W^{d}( -\bl_2, \bl_2' )
\\
\quad\quad\quad +
C_{| \bl_1 - \bl_2' |}^{cd} 
W^{c}( -\bl_1,  \bl_2' )
W^{d}( -\bl_2, \bl_1' )
\Big\}
\\ \times F_{ | \bl_1 |} F_{ | \bl_2 | } 
W^{x}( \bl_1,  \bl_2)
W^{z}( \bl_1', \bl_2'),
\end{multline}
where the trispectrum response function is defined as
\be
\left. {R}^{xz}_{L, ab} \right|_{cd} \equiv 
\frac{ 
	{\cal R}^{xc}_L {\cal R}^{zd}_L + {\cal R}^{xd}_L {\cal R}^{zc}_L 
	}{ 
	{\cal R}^{xa}_L {\cal R}^{zb}_L + {\cal R}^{xb}_L {\cal R}^{za}_L 
	}.
\ee

\end{document}